\documentclass[prd,twocolumn,showpacs,showkeys,amsmath,amssymb,tightenlines]{revtex4}
\usepackage{graphicx}
\usepackage[bookmarksopen]{hyperref}

\def  \o    {\omega}

\def  \s    {\sigma}

\def  \p    {\pi}

\def  \m    {\mu}
\def  \n    {\nu}

\def  \f    {\frac}
\def  \lt   {\left}
\def  \rt   {\right}
\def  \th   {\theta}

\def  \ra   {\rightarrow}

\def  \veps {\varepsilon}

\def  \gm   {\gamma^\mu}

\def  \del  {\partial}
\def  \cl   {{\cal{L}}}

\def  \mn   {\mu \nu}

\def  \bef  {\begin{figure}}
\def  \eef  {\end{figure}}
\def  \be   {\begin{equation}}
\def  \ee   {\end{equation}}
\def  \ba   {\begin{array}}
\def  \ea   {\end{array}}
\def  \bea  {\begin{eqnarray}}
\def  \eea  {\end{eqnarray}}
\def  \beq  {\begin{eqnarray}}
\def  \eeq  {\end{eqnarray}}
\def  \nn   {\nonumber}
\def  \bd   {\begin{displaymath}}
\def  \ed   {\end{displaymath}}
\def  \bse  {\begin{subequations}}
\def  \ese  {\end{subequations}}
\def  \bwt  {\begin{widetext}}
\def  \ewt  {\end{widetext}}

\def  \ba   {{\bf{a_1}}}

\begin{document}

%\begin{frontmatter}

\title{Pionic contribution to relativistic Fermi Liquid parameters}

\author {Kausik Pal}
\email {kausik.pal@saha.ac.in}
\affiliation{High Energy Physics  Division, Saha Institute of Nuclear Physics, 
1/AF Bidhannagar, Kolkata 700064, INDIA.}

\medskip
\begin{abstract}
We calculate pionic contribution to the relativistic Fermi Liquid parameters 
(RFLPs) using Chiral Effective Lagrangian. The RFLPs so 
determined are then used to calculate chemical potential, exchange and nuclear
symmetry energies due to $\pi$$N$ interaction. We also evaluate 
two loop ring diagrams involving
$\sigma$, $\omega$ and $\pi$ meson exchanges
and compare results with what one obtains from the
relativistic Fermi Liquid theory (RFLT).
\end{abstract}
\vspace{0.08 cm}

\keywords {Pion, Fermi Liquid Parameters, Exchange energy.}
\pacs{21.65.-f, 13.75.Cs, 13.75.Gx, 21.30.Fe}

\maketitle

\section{Introduction}
Fermi liquid theory (FLT) provides us with one of the most important
theoretical scheme to study the properties of strongly interacting Fermi
systems involving low lying excitations near the Fermi surface {\cite{song01}}.
Although developed originally in the context of studying the properties of 
$^3He$, it  has  widespread applications in other disciplines of many body
physics like superconductivity, super fluidity, nuclear and
neutron star matter etc {\cite{baym_book,chin76}}. 

In nuclear physics FLT was first extended and used by
Migdal \cite{migdal78} to study the properties of both unbound 
nuclear matter and Finite nuclei \cite{mig_book}. FLT also provides theoretical
foundation for the nuclear shell model \cite{mig_book} as well as nuclear 
dynamics of low energy excitations\cite{baym_book,krewald88}. Particularly, 
ref.\cite{brown71} reveals the connection between Landau, Brueckner-Bethe and
Migdal theories, ref.{\cite{celenza_book,anastasio83}}, on the other
hand, calculates 
the Migdal parameters using one-boson-exchange models of the nuclear 
force and shows how these parameters 
are modified if nuclear matter is considered within the context of Relativistic
Brueckner-Hartree-Fock (RBHF) model. Some of the recent work that examines
Fermi-liquid properties of hadronic matter also incorporates Brown-Rho (BR) 
scaling which is very important for the study of the properties of hadrons
in dense nuclear matter (DNM){\cite{friman96,friman99}}.

Most of the earlier nuclear matter calculations that involved Landau theory
were done in a non-relativistic framework.
The relativistic extension of the FLT was first developed by Baym and Chin 
{\cite{chin76}} in the context of studying the properties of DNM. 
In \cite{chin76} the authors invoked Walecka model (WM) to calculate various  interaction parameters ($f_{pp^\prime}$) but did not consider
mean fields (MF) for the $\s$ and $\omega$ meson {\em i.e.} there the FLPs are calculated perturbatively.

Later Matsui revisited the 
problem in \cite{matsui81} where one starts from the expression of energy 
density in presence of scalar and vector meson MF and
takes functional derivatives to determine the FLPs. The results are 
qualitatively different than the perturbative results 
as may be seen from \cite{chin76,matsui81}. A comparison of relativistic and
non-relativistic calculations have been made in 
\cite{celenza_book,anastasio83} which
also discusses how the FLPs are modified in presence of the $\s$ and
$\o$ MF and contrast those with perturbative results.

Besides $\s$ and $\o$ meson, ref.\cite{matsui81} 
also includes the $\rho$ and $\p$ meson and the model adopted was originally
proposed  by Serot to incorporate pion into the WM. 
It is to be noted, however, that the  FLPs presented in \cite{matsui81} are
independent of $\p$ meson. This is because $\p$, being a pseudoscalar,
fails to contribute at the MF level. Hence to estimate the pionic
contribution to FLPs it is necessary to go beyond MF formalism.
 
It is to be noted that, to our knowledge, such relativistic
calculations including $\p$ exchange does not exist, 
despite the fact that pion has a 
special status in nuclear physics as it is responsible for 
the spin-isospin dependent long range nuclear force. Furthermore,
there are various non-relativistic calculations including the 
celebrated work of Migdal which shows that the pionic contribution to FLPs are 
important {\cite{friman96,friman99,rho80}} and most dominant for low energy
excitations. It might be mentioned
here that  \cite{friman99} discusses how can one incorporate
relativistic corrections to $F_1^\pi$ in the static potential model 
calculation. We, however, take the approach of \cite{chin76} where all 
the fields are treated relativistically.

The other major departure of the present work from ref.\cite{chin76,matsui81} 
resides in the choice of model for the description of the many body nuclear 
system. Unlike previous calculations \cite{chin76,matsui81}, here 
we use recently developed Chiral Effective Field Theory (chEFT) to
calculate the RFLPs. Such a choice is motivated by the following earlier
works that we now briefly review. 

The first attempt to include $\pi$ and $\rho$ meson into the WM was made
in \cite{serot79} where to describe the $\pi$N dynamics pseudoscalar (PS)
coupling was used. Such straightforward inclusion of $\pi$ meson into
the WM has serious difficulties. In particular, at the MF level it gives 
rise to tachyonic mode  
in matter at densities as low
as $ 0.1\rho_0$, where $\rho_0$ is the nuclear saturation
density \cite{kapusta81}. 
Inclusion of exchange diagrams removes
such unphysical mode but makes the effective mass unrealistically large.
If one simply replaces the PS coupling of ref.\cite{serot79} with the 
pseudovector (PV) interaction, these difficulties can be avoided. This, 
however, turns the theory non-renormalizable
\cite{serot82}. 

In \cite{serot82} it was shown how, starting from a PS coupling one can arrive
at PV representation which preserves renormalizibility of the theory  and 
at the same time yields realistic results for the pion dispersion relations 
in matter. But this model was also
not found to be trouble free, particularly it had several
shortcomings in describing $\pi$$N$ dynamics in matter which we do not
discuss here and refer the reader to {\cite{furnstahl87,furnstahl93,furnstahl95,
furnstahl96,serot97,biswas08}}. 

Furthermore, the WM itself has several 
problems in relation to convergence which forbids systematic expansion scheme 
to perform any perturbative calculations. This was first exposed in 
ref.\cite{furnstahl89}.

The most recent model that provides us with a systematic
scheme to study the dense nuclear system is provided by chEFT 
{\cite {serot97,furnstahl89,furnstahl97,hu07}} 
where 
the criterion of renormalizibility is
given up in favor of successful relativistic description of DNM and the
properties of finite nuclei.
The chEFT, apart from $\s$ and $\o$ mesons, 
also includes pion and therefore is best suited for the present purpose.

In this paper we estimate contributions of pion exchange to the 
FLPs within the framework of RFLT and subsequently use the parameters
so determined to calculate various quantities like pionic contribution
to the chemical potential, energy density, symmetry energy ($\beta$) etc. 
For completeness and direct comparison with the two loop results we also 
calculate here the exchange energy due to the interaction mediated by the 
$\s$, $\o$ and $\p$ mesons.

\section{Formalism}

Now we quickly outline the formalism.
In FLT the energy density $E$ of an interacting system is the functional of 
occupation number $n_{p}$ of the quasi-particle states of momentum $p$. 
The excitation of the system is equivalent to the change of occupation 
number by an amount $\delta n_{p}$. The corresponding energy of the system is 
given by {\cite{baym_book,chin76}},
\beq{\label {total_energy}}
E&=&E^{0}+\sum_{s}\int\frac{d^3{p}}{(2\pi)^3}
\varepsilon_{ps}^{0}\delta n_{ps}
\nn\\&&
+\frac{1}{2}\sum_{ss'}\int\frac{d^3{p}}{(2\pi)^3}\frac{d^3{p'}}{(2\pi)^3}
f_{ps,p's'}
\delta n_{ps}\delta n_{p's'}.
\eeq 
Here, $s$ is the spin index. The quasi-particle energy can be 
written as,
\beq\label{quasi_energy}
\veps_{ps}=\veps_{ps}^{0}+\sum_{s'}\int\frac{d^3{p'}}{(2\pi)^3}f_{ps,p's'}
\delta n_{p's'},
\eeq
where superscript $0$ denotes the ground state \cite{baym_book}.
It is to be remembered, that, 
although the interaction ($f_{ps,p's'}$) between the 
quasiparticles is not small, the problem is greatly simplified because 
it is sufficient to consider only pair collisions between the quasiparticles
{\cite{mig_book}}.

Since quasi-particles are well defined only near the Fermi surface, one assumes
\beq
\left.\begin{array}{lll} 
&\veps_{p}&=\mu+v_{f}(p-p_{f})\\ 
{\rm and~~~} & p&\simeq p'\simeq p_{f}.
\end{array}
\right\}
\eeq
 Then LPs $f_{l}$s are defined by the Legendre expansion of $f_{ps,p's'}$ as
{\cite{baym_book,chin76}},
\beq\label{landau_para}
f_l=\frac{2l+1}{4}\sum_{ss'}\int\frac{d\Omega}{4\pi}P_{l}(\cos\theta)f_{ps,p's'}
\eeq
where $\theta$ is the angle between $p$ and $p'$, both taken to be on the Fermi 
surface, and the integration is over all directions of $p$. We
 restrict ourselves for $l\le 1$ {\em i.e.} $f_{0}$ and $f_{1}$, as higher $l$ 
contribution decreases rapidly {\cite{chin76,matsui81}}.

Now the Landau Fermi liquid interaction $f_{ps,p's'}$ is related to the
 two particle forward scattering amplitude via {\cite{chin76}},
\beq
f_{ps,p's'}&=&\frac{M}{\veps_{p}^0}\frac{M}{\veps_{p'}^0}
{\cal M}_{ps,p's'},
\eeq
 where the Lorentz invariant matrix ${\cal M}_{ps,p's'}$ consists
of the usual direct and exchange amplitude, which may be evaluated directly
from the relevant Feynman diagrams.
 The spin averaged scattering amplitude  $(f_{pp'})$ is given by 
{\cite{chin76}},
\beq\label{fermi_inter}
f_{pp'}&=&\frac{1}{4}\sum_{ss'}\frac{M}{\veps_{p}^0}
\frac{M}{\veps_{p'}^0}{\cal M}_{ps,p's'}.
\eeq 

The dimensionless LPs are 
$F_{l}=N(0)f_{l}$, where $N(0)$ is the density of states at the Fermi surface 
defined as {\cite{matsui81}},
\beq\label{dens_of_state} 
N(0)&=&\frac{g_{s}g_{I}p_{f}\veps_{f}}{2\pi^2}.
\eeq
 Here $g_{s},g_{I}$ are 
the spin and isospin degeneracy factor respectively.

\section{Landau Parameters}

By retaining only the lowest order terms in the pion fields, one obtains the 
following Lagrangian from the chirally invariant Lagrangian 
{\cite{furnstahl97,hu07}}:
\beq
\cl &=& \bar{\Psi}\left[\gm(i\del_{\mu}-g_{\o}\o_{\mu})-i\frac{g_{A}}{f_{\pi}}
\gm\gamma_{5}\del_{\mu}\underline{\pi}-(M-g_{\s}\Phi_{\s})\right]\Psi\nn\\&&
+\frac{1}{2}\del^{\mu}\Phi_{\s}\del_{\mu}\Phi_{\s}-\frac{1}{2}m_{\s}^2\Phi_{\s}^2
-\frac{1}{4}\o^{\mn}\o_{\mn}+\frac{1}{2}m_{\o}^2\o^{\m}\o_{\m}\nn\\&&
+\frac{1}{2}\del^{\m}{\vec\Phi_{\p}}\cdot\del_{\m}{\vec\Phi_{\p}}
-\frac{1}{2}m_{\p}^2{\vec\Phi_{\p}^2}
+\cl_{NL}+\delta\cl,
\eeq
where $\o_{\mn}=\del_{\m}\o_{\n}-\del_{\n}\o_{\m}$,
$\underline{\p}=\frac{1}{2}({\vec\tau}\cdot{\vec\Phi_{\p}})$ and 
${\vec\tau}$ is the isospin index. Here $\Psi$ is the nucleon
field and $\Phi_{\s}$, $\o_{\m}$ and $\Phi_{\p}$ are the meson fields 
(isoscalar-scalar, isoscalar-vector and isovector-pseudoscalar respectively).
The terms $\cl_{NL}$ and $\delta\cl$ contain the non-linear  
and counterterms respectively (for explicit expressions  
see {\cite{hu07}}).

Now due to presence of pion fields in the chiral Lagrangian we have 
component in 
the interaction which acts on the isospin fluctuation. One can derive the 
isospin dependent quasiparticle interaction along the line of 
ref.{\cite{chin76}}. For pions, as mentioned before,
the direct term vanishes and it is only the 
exchange diagram that contributes to the interaction parameter :
\beq\label{pion_interaction} 
f_{pp'}^{ex,\p}{\bigg\vert}_{p\simeq p'=p_{f}} & =& \frac{3}{2}
\frac{g_{A}^2M^{*2}}{4f_{\p}^2\veps_{f}^2} 
\lt\{\f{p_{f}^2(1-\cos\th)}{2p_{f}^2(1-\cos\th)+m_{\p}^2}\rt\},
\eeq

where $\veps_{f}=(p_{f}^2+M^{*^2})^{1/2}$ and $g_{A}^2=1.5876$ , $f_{\p}=93 MeV$, 
$m_{\p}=139$ MeV{\cite{hu07}}.
The factor $3/2$ arises since the isospin factor is $3/2$ in the isoscalar channel {\cite{friman99}}.

The effective nucleon mass $M^*$ is 
determined self-consistently from the following equation {\cite {matsui81}},
\beq\label{eff_M}
M^*&=&M-\frac{g_{\s}^2}{m_{\s}^2}
{\sum_{i}}n_{i}\frac{M^*}{(p_{i}^2+M^{*2})^{1/2}}.
\eeq
Using Eq.(\ref{landau_para}) and Eq.(\ref{pion_interaction}) we can derive 
isoscalar LPs $f_{0}^{ex,\p}$ and $f_{1}^{ex,\p}$, 
\beq\label{f0_pi}
f^{ex,\p}_{0}&=&-\f{3g_{A}^2M^{*2}}{32f_{\p}^2\veps_{f}^2}
\left[-2+\frac{m_{\pi}^2}{2p_{f}^2}
\ln\left(1+\frac{4p_{f}^2}{m_{\pi}^2}\right)\right],
\eeq
and
\beq\label{f1_pi}
\frac{1}{3}f^{ex,\pi}_{1}&=&-\f{3g_{A}^2M^{*2}m_{\p}^2}
{64f_{\p}^2\veps_{f}^2p_{f}^2}\times\nn\\&&
\left[-2+\left(\frac{m_{\pi}^2}
{2p_{f}^2}+1\right)\ln\left(1+\frac{4p_{f}^2}{m_{\pi}^2}\right)\right].
\eeq

Using Eqs.(\ref{f0_pi}) and (\ref{f1_pi}) we find that
\beq\label{f0f1pi}
f^{ex,\p}_{0}-\f{1}{3}f^{ex,\p}_{1}
=\f{3g_{A}^2M^{*2}}{16f_{\p}^2\veps_{f}^2}\times\nn\\
\left[\frac{m_{\pi}^4}
{8p_{f}^4}\ln\left(1+\frac{4p_{f}^2}{m_{\pi}^2}\right)
-\f{m_{\p}^2}{2p_{f}^2}+1\right].
\eeq
It is this combination {\em i.e.} $f_{0}-\frac{1}{3}f_{1}$, which appears in 
the calculation of chemical potential and other relevant quantities.
For the massless pion, Eq.(\ref{f0f1pi}) turns out to be finite,
\beq\label{f0f1pi1}
\left(f^{ex,\p}_{0}-\f{1}{3}f^{ex,\p}_{1}\right){\Bigg\vert}_{m_{\p}\ra 0}
&=&\f{3g_{A}^2M^{*2}}{16f_{\p}^2\veps_{f}^2}.
\eeq

It is to be noted that, in the massless limit
for $\sigma$ and $\omega$ meson, 
$f^{ex}_{0}$ and $f^{ex}_{1}$ diverge as shown in \cite{chin76}, in contrast
for  pion, even in the massless limit, these are finite. This is due to the 
presence of $(1-\cos\theta)$ in the numerator 
of Eq.(\ref{pion_interaction}) unlike $\s$ and $\o$ meson. 

The dimensionless LPs can be determined
by equating the equation $F_{0}=N(0)f_{0}$ and 
$F_{1}=N(0)f_{1}$, where
 $N(0)$ is the density of states at the Fermi surface defined in 
Eq.(\ref{dens_of_state}). Thus the dimensionless parameters are
\beq\label{dless_F0}
F^{ex,\p}_{0}=-g_{s}g_{I}\f{3g_{A}^2p_{f}M^{*2}}{64\p^2f_{\p}^2\veps_{f}}
\lt[-2+\frac{m_{\pi}^2}{2p_{f}^2}
\ln\left(1+\frac{4p_{f}^2}{m_{\pi}^2}\right)\rt],
\eeq
and
\beq\label{dless_F1}
\f{1}{3}F^{ex,\p}_{1}=-g_{s}g_{I}\f{3g_{A}^2m_{\p}^2M^{*2}}
{128\p^2f_{\p}^2p_{f}\veps_{f}}\times\nn\\
\lt[-2+\left(\frac{m_{\pi}^2}
{2p_{f}^2}+1\right)\ln\left(1+\frac{4p_{f}^2}{m_{\pi}^2}\right)\rt].
\eeq

\vskip 0.2in
%%%%%%%%%%%%%%%%%%%%%%%%%%%%%%%%%%%%%%%%%%%%%%%%%%%%%%%%%%%%%%%%%%%%
%%%%%%%%%%%%%%%%%%%%        FIG - 1             %%%%%%%%%%%%%%%%%%%
%%%%%%%%%%%%%%%%%%%%%%%%%%%%%%%%%%%%%%%%%%%%%%%%%%%%%%%%%%%%%%%%%%%%

\begin{figure}
%\begin{center}
\resizebox{7cm}{5.0cm}{\includegraphics[]{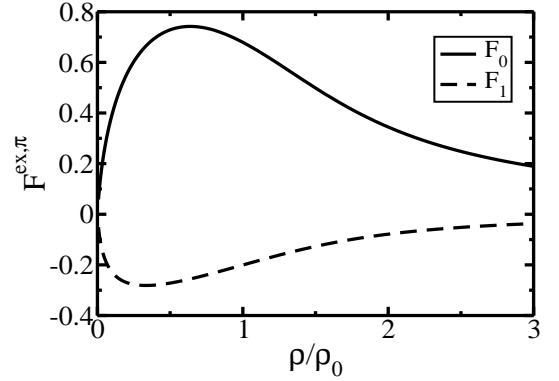}}
\caption{Dimensionless Landau parameters in symmetric nuclear 
matter for pion exchange in relativistic theory. 
Solid and dashed line represent $F_{0}$ and $F_{1}$ respectively.}
\label{fig1}
%\end{center}
\end{figure}

%%%%%%%%%%%%%%%%%%%%%%%%%%%%%%%%%%%%%%%%%%%%%%%%%%%%%%%%%%%%%%%%%%%%%

In Fig.(\ref{fig1}) we show the density dependent of $F_{0}$ and 
$F_{1}$ due to pionic interaction. Numerically 
at nuclear matter density $(\rho_{0}=0.148 {\rm fm^{-3}})$, 
$F^{ex,\p}_{0}=0.68$ and 
$F^{ex,\p}_{1}=-0.2$. In the non-relativistic limit $\veps_{f}\rightarrow M^{*}$, one obtains the same expression for $F_{1}^{\pi}$ as reported in 
\cite{{friman96},{rho80}}.

\section{Chemical potential}

We now proceed to calculate the  chemical potential which in principle will
be different for the neutron and proton in asymmetric nuclear matter.

Although in the present work we deal with symmetric nuclear matter (SNM) for which 
$\mu_p=\mu_n$. To calculate the general expression for chemical potential with arbitrary asymmetry $(\alpha=(n_n-n_p)/(n_n+n_p))$, we take distribution function with explicit isospin index $(b ~or~ b')$, so that variation of distribution function gives
{\cite{baym_book}}
\beq\label{delta_nb}
\delta n_{b}&=& -N_{b}(0)\left[\sum_{b'}f_{0}^{bb'}\delta n_{b'}-\delta\mu_{b}\right],
\eeq 
where $N_{b}(0)$ is the isospin dependent density of states at the corresponding Fermi surface. The Eq.(\ref{delta_nb}) yields
\beq
\frac{\del\mu_{b}}{\del n_{b}}&=&\frac{1}{N_{b}(0)}+\sum_{b'}f_{0}^{bb'}
\frac{\del n_{b'}}{\del n_{b}}.
\eeq
In our case $b~(or~ b')= n,p$. Separately for neutron and proton we have
\beq\label{general_mu}
\left(\begin{array}{c}
\del{\mu_n}\\
\del{\mu_p}
\end{array}\right)
&=&\left(\begin{array}{cc}
\frac{1}{N_n(0)}+f_{0}^{nn}  &  f_{0}^{np}\\
     f_{0}^{pn}              &  \frac{1}{N_p(0)}+f_{0}^{pp}
\end{array}\right)
\left(\begin{array}{c}
\del{n_n}\\
\del{n_p}
\end{array}\right),\nn\\
\eeq
where the superscripts $nn$ and $np$ denote neutron-neutron and neutron-proton quasiparticle interaction. $N_n(0)$ and $N_p(0)$ denote the density of states at  neutron and proton Fermi surfaces respectively {\cite{aguirre07, sjoberg76}}. For pion exchange, in SNM, $f_{l}^{np(pn)}=2f_{l}^{nn(pp)}$ 
{\cite{rho80}}, from Eq.(\ref{general_mu}), we have,
\beq
\del \mu_{n}&=&\frac{1+3F_{0}^{nn}}{N_n(0)}\del n_{n},
\eeq
where $F_{0}^{nn}=N_{n}(0)f_{0}^{nn}$.
Similarly, one can determine $\del \mu_{p}$.  
Motivated by {\cite{mig_book}}, we define
\beq
f_{l}^{b}&=&\frac{1}{2}\sum_{b'}f_{l}^{bb'}.
\eeq 
Evidently in SNM, $f_{l}^{nn}=f_{l}^{pp}$, and therefore, we write 
$\mu_{n}=\mu_{p}=\mu$ {\cite{chin76}},
\beq\label{mudmu}
\mu{\rm d\mu}=\left[p_{f}+g_{s}g_{I}\frac{\mu p_{f}}{2\pi^2}
\left(f_{0}-\frac{1}{3}f_{1}\right)\right]dp_{f}.
\eeq
Here $f_{0}$ and $f_{1}$ are given by Eq.(\ref{f0_pi}) and (\ref{f1_pi}).

To calculate $\mu$, it is sufficient to let $\mu=\veps_{f}$ in the right hand 
side of Eq.(\ref{mudmu}). With the constant of integration adjusted so that
at high density $ p_{f}\simeq \veps_{f}$, Eq.(\ref{mudmu}) upon integration 
together with Eq.(\ref{f0f1pi}) yield
\beq
\mu &=&\veps_{f}-g_{s}\f{3g_{A}^2M^{*4}}{128\p^2f_{\p}^2\veps_{f}}
\times\nn\\&&
\left[-2y_{\pi}^3
\sqrt{-y_{\pi}^2+4}\tan^{-1}\left(\frac{x\sqrt{-y_{\pi}^2+4}}{y_{\pi}\sqrt{1+x^2}}
\right)\right.\nn\\&&\left.
+\frac{y_{\pi}^4\sqrt{1+x^2}}{x}\ln\left(1+\frac{4x^2}{y_{\pi}^2}\right)
-4x\sqrt{1+x^2}\right.\nn\\&&\left.
-2(y_{\pi}^4-2y_{\pi}^2-2)\ln(x+\sqrt{1+x^2})\right],
\eeq
where $x=p_{f}/M^*$ and $y_{\p}=m_{\p}/M^*$.

The calculations of LPs for other mesons is
straightforward. However, for brevity, we
do not present corresponding expressions for $\s$ and $\o$
mesons but quote their numerical values in Table(\ref{table-1}).
The numbers cited above are relevant for normal nuclear matter
density $\rho_{0}=0.148 {\rm fm^{-3}}$. For the coupling constants we 
adopt the same parameter set as designated by {\bf M0A} in {\cite{hu07}}.

Interestingly, individual contribution to LPs of $\s$ and $\o$ meson are
large while sum of their contribution to $F_{0}^{tot}$ is small due to the
sensitive cancellation of $F_{0}^{\s}$ and $F_{0}^{\o}$ as can be seen
from Table(\ref{table-1}). 
Such a cancellation is responsible for the nuclear saturation dynamics  
{\cite{celenza_book,anastasio83}}.  Numerically, $F_{0}^{\s+\o}$ is approximately $3/2$ times smaller than $F_{0}^{\pi}$. 

%%%%%%%%%%%%%%%%%%%%%%%%%%%%%%%%%%%%%%%%%%%%%%%%%%%%%%%%%%%%%%%%%%%%
%%%%%%%%%              TABLE-1             %%%%%%%%%%%%%%%%%%%%%%%%
%%%%%%%%%%%%%%%%%%%%%%%%%%%%%%%%%%%%%%%%%%%%%%%%%%%%%%%%%%%%%%%%%%%%%

\begin{table}
\caption{Dimensionless Landau parameters and chemical potential at 
$\rho=\rho_{0}$. Note that, $F_{0}$, $F_{1}$ are the dimensionless 
isoscalar LPs.}
\label{table-1}
\begin{center}
\begin{tabular}{ccccc} \hline \hline
 Meson   & $F_{0}$ & ~~~~~~ & $F_{1}$ \\ \hline
  $\s$   &  -5.04  & ~~~~~~ & 0.875   \\
  $\o$   &   5.44  & ~~~~~~ & -0.93    \\ 
  $\p$   &   0.68   & ~~~~~~ & -0.20   \\ \hline\hline
\end{tabular}
\end{center}
\end{table}

%%%%%%%%%%%%%%%%%%%%%%%%%%%%%%%%%%%%%%%%%%%%%%%%%%%%%%%%%%%%%%%%%%%%%%

\section{Exchange energy}

Once the $\mu $ is determined, one can readily calculate the energy density due 
to $\pi-N$ interaction in SNM as {\cite{chin76,chin77}},
\beq\label{xe_pi}
E^{ex}_{\p}&=&\int{\rm d\rho}(\mu -\veps_{f})\nn\\
&=&-g_{s}^2\frac{3g_{A}^2M^{*6}}{128f_{\p}^2\p^4}\times\nn\\&&
\left[I_{\p}+y_{\p}^4\left\{-\frac{x^2}{2}+\frac{y_{\p}^2+4x^2}{8}
\ln\left(1+\f{4x^2}{y_{\p}^2}\right)\right\}-x^4\right.\nn\\
&&\left.
+\left(\f{y_{\p}^4}{2}-y_{\p}^2-1\right)
\left(\{x\eta-\ln(x+\eta)\}^2-x^4\right)\right],
\eeq
where $\eta=\sqrt{1+x^2}$ and 
\beq
I_{\p}=-2y_{\p}^3\sqrt{4-y_{\p}^2}
\int\f{x^2}{\eta}
\tan^{-1}\left(\frac{x\sqrt{4-y_{\p}^2}}{y_{\p}\eta}\right)
{\rm d}x.
\eeq
For the massless pion this reads as
\beq\label{e_mless}
E^{ex}_{\p}{\Big\vert}_{m_{\p}\ra 0}&=&\f{3g_{A}^2M^{*6}}{32f_{\p}^2\p^4}
[x\eta-\ln(x+\eta)]^2.
\eeq

\vskip 0.2in
%%%%%%%%%%%%%%%%%%%%%%%%%%%%%%%%%%%%%%%%%%%%%%%%%%%%%%%%%%%%%%%%%%%%%%%%
%%%%%%%%%%%%%%%%           FIG - TWO_LOOP_JAX             %%%% %%%%%%%%%
%%%%%%%%%%%%%%%%%%%%%%%%%%%%%%%%%%%%%%%%%%%%%%%%%%%%%%%%%%%%%%%%%%%%%%%%

\begin{figure}
%\begin{center}
\resizebox{8cm}{3.0cm}{\includegraphics[]{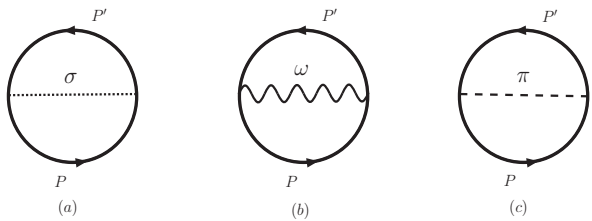}}
\caption{Two-loop contributions to the nuclear matter energy density. 
The solid line represents the baryon propagator. $\sigma$, $\omega$
and $\pi$ mesons are denoted by dotted, wavy and dashed line respectively.}
\label{two_loop_jax}
%\end{center}
\end{figure}

%%%%%%%%%%%%%%%%%%%%%%%%%%%%%%%%%%%%%%%%%%%%%%%%%%%%%%%%%%%

The contribution arising from pion exchange from the direct evaluation 
of Fig.\ref{two_loop_jax}(c) reads as {\cite{hu07}},
\beq\label{ring_eng}
E^{ex}_{\p}=g_{s}g_{I}\frac{3g_{A}^2}{32\p^4f_{\p}^2}M^{*2}
\int_{0}^{p_{f}}\frac{|p|^2{\rm dp}}{\veps_{p}}
\int_{0}^{p_{f}}\frac{|p'|^2{\rm dp'}}{\veps_{p'}}
\times\nn \\ \int_{-1}^{1}{\rm d(\cos\theta)}
\left[\frac{\veps_{p}\veps_{p'}-|p||p'|\cos\theta-M^{*2}}
{2\veps_{p}\veps_{p'}-2|p||p'|\cos\theta-2M^{*2}+m_{\p}^2}\right].
\eeq
Similarly, from Fig.\ref{two_loop_jax}-((a),(b)) one can determine exchange 
energy due to $\s$ and $\o$ meson interaction {\cite{hu07}}. 
In Table(\ref{table-2}) we given the exchange energy results calculated 
from RFLPs which is in agreement with two-loop results {\cite{hu07}}. 
It might be mentioned here that in the limit $m_{\pi}\rightarrow 0$, the 
Eq.(\ref{ring_eng}) can be evaluated analytically which reproduces the 
Eq.(\ref{e_mless}) derived in FLT approach.

%\vskip 0.2in

%%%%%%%%%%%%%%%%%%%%%%%%%%%%%%%%%%%%%%%%%%%%%%%%%%%%%%%%%%%%%%%%%%%%
%%%%%%%%%              TABLE-2            %%%%%%%%%%%%%%%%%%%%%%%%
%%%%%%%%%%%%%%%%%%%%%%%%%%%%%%%%%%%%%%%%%%%%%%%%%%%%%%%%%%%%%%%%%%%%%

\begin{table}
\caption{Exchange energy in MeV from FLT at $\rho=\rho_{0}$.}
\begin{center}
\label{table-2}
\begin{tabular}{ccc} \hline \hline
 Meson    &~~~~     & $E^{ex}$    \\ 
          &~~~~     &  FLT        \\  \hline
  $\s$    &~~~~     &  40.48      \\
  $\o$    &~~~~     & -23.41       \\ 
  $\p$    &~~~~     &  12.49       \\    \hline\hline
\end{tabular}
\end{center}
\end{table}

%%%%%%%%%%%%%%%%%%%%%%%%%%%%%%%%%%%%%%%%%%%%%%%%%%%%%%%%%%%%%%%%%%%%%%

\section{Symmetry energy}

Knowing the "isovector" combination of the LPs, one can determine nuclear symmetry energy. The symmetry
energy is defined as the difference of energy between the neutron matter and
symmetric nuclear matter is given by the following expression 
{\cite{matsui81,greco03}}
\beq
\beta&=&\frac{1}{2}\rho\frac{\del^{2}E}{\del\rho_{3}^2}
{\bigg\vert}_{\rho_{3}=0}.
\eeq
In terms of LPs, the symmetry energy can be expressed as {\cite{mig_book}}
\beq\label{symm_energy}
\beta = \frac{p_{f}^2}{6\veps_{f}}(1+2F^{\prime}_{0}),
\eeq
where $F_{0}^{\prime}=\frac{1}{2}(F_{0}^{nn}-F_{0}^{np})$ is the isovector combination of dimensionless Landau parameters $F_{0}$ {\cite{mig_book,greco03}}. Deriving $F_{0}^{\prime}$, one can find symmetry 
energy. Numerically at saturation density ($\rho=\rho_{0}$) we obtain 
$\beta = 14.57$ MeV. So relatively small contribution to $\beta$ comes from 
one pion exchange diagram \cite{dieper03}

\section{Summary and conclusion}

In this paper, we calculate RFLPs within the framework of
RFLT. For the description of dense
nuclear system chEFT is invoked. 
Although our main focus was to estimate the 
contribution of pions to RFLPs, for comparison and completeness we also
present results for the $\s$ and $\o$ meson. 
It is seen 
that the pionic contribution to the FLPs are significantly larger 
compared to the combined contributions of $\s$ and $\o$ meson. 
Thus any realistic relativistic calculation for the FLPs
should include $\p$ meson which necessarily implies going beyond the
MF calculations. 
The LPs what we determine here 
are subsequently used to calculate exchange and symmetry energy of the system.
Finally we evaluate two loop ring diagrams with the same 
set of interaction parameters 
and show that the numerical results are consistent with those obtained from
the FLT. It might be mentioned here that in the present calculation we 
have ignored nucleon-nucleon correlations which might be worthwhile
to investigate. It should, however, be noted that inclusion of correlation energy would require the readjustment of the coupling parameter so as to reproduce the saturation properties of nuclear matter. 

\vskip 0.2in
{\bf Acknowledgments}\\

The author would like to thank G.Baym and C.Gale for their valuable comments. 
I also wish to thank A.K.Dutt-Mazumder 
for the critical reading of the manuscript.

%%%%%%%%%%%%%%%%%%%%%%%%%%%%%%%%%%%%%%%%%%%%%%%%%%%%%%%%%%%%%
%%%%%%%%%%%%          REFERENCES         %%%%%%%%%%%%%%%%%%%
%%%%%%%%%%%%%%%%%%%%%%%%%%%%%%%%%%%%%%%%%%%%%%%%%%%%%%%%%%%%%

\end{document}